\documentclass[manuscript]{revtex4-1}
\usepackage{amsmath,amssymb,graphicx,subfig,upgreek,siunitx}
\usepackage{graphicx}
\usepackage{color}
\newcommand{\om}{\omega}

\begin{document}

\title{{Analogy between thermal emission of nano objects and Hawking's radiation}}
\author{Karl Joulain}
\affiliation{Institut Pprime, CNRS, Universit\'e de Poitiers, ISAE-ENSMA, F-86962 Futuroscope Chasseneuil, France}

\date{\today}
\begin{abstract}
We analyze in this work some analogies between thermal emission of nano objects and Hawking's radiation. We first focus on the famous expression of the black hole radiating temperature derived by Hawking in 1974 and consider the case of thermal emission of a small aperture made into a cavity (Ideal Blackbody). We show that an expression very similar to Hawking's temperature determines a temperature below which an aperture in a cavity cannot be considered as standard blackbody radiating like $T^4$. Hawking's radiation therefore appear as a radiation at a typical wavelength which is of the size of the horizon radius. In a second part, we make the analogy between the emission of particle-anti particle pairs near the black hole horizon and the scattering and coupling of thermally populated evanescent waves by a nano objects. We show here again that a temperature similar to the Hawking temperature determines the regimes where the scattering occur or where it is negligible.

\end{abstract}

\maketitle

\section{Introduction. Hawking temperature.}
In 1974, Stephen Hawking \cite{hawking_black_1974,hawking_particle_1975} published his famous work about black hole radiating temperature. This famous temperature expression is the sole one in physics in which appears all 4 fundamentals constants ($c$, $h$, $G$ and $k_b$). The origin of this radiation lies in the fact that the black hole makes vacuum fluctuations such as particle-antiparticle pairs radiate at infinity since they are separated at the black hole horizon : a particle is radiated to infinity whereas its antiparticle is pursuing its path into the black hole. The idea of the present work is two fold. First, we will remark that a similar formula for a critical temperature is obtained when one is studying thermal emission of an aperture. Second, an analogy is made between the particle-antiparticle pair separation at the horizon and the scattering of evanescent waves such as polariton by subwavelength structures. It is shown in this second part, that here again, this coupling to the far-field of evanescent waves thermally populated happens when the temperature is smaller than a critical temperature very similar to the Hawking one.

Hawking temperature formula can be heuristically derived  from simple principles like the Heisenberg incertitude relation as well as simple relations between distance and velocity or wavelength and frequency. The idea is that in vacuum, there are quantum fluctuations that can be interpreted as permanent creation and annihilation of particle-antiparticle pairs. These pairs remain virtual as long as the Heisenberg incertitude relations are violated i.e the energy fluctuation $\Delta E$ due to the vacuum fluctuations and particle antiparticle pair formation multiplied by the time of their existence $\Delta t$ is smaller than the Planck constant ($\Delta E\Delta t\leq \hbar$). In the presence of a black hole horizon, the particle-antiparticle pairs close to the horizon can be separated. The typical time $\Delta t$ on which the pair are separated is the order of the time required to cross the whole horizon black hole that is  $\Delta t\sim 2 \pi R_s/c$ where $R_s=2GM/c^2$ is the Schwarzschild radius. Using the fact that the virtual pairs will be seen when the Heisenberg incertitude relations are valid, a particle that will be seen at infinite coming from the black hole will have their energy $\Delta E$ that will obey
$$
\Delta E\Delta t\sim\hbar/2
$$ 
that is
$$
\Delta E\sim\frac{hc}{8\pi^2R_s}
$$
Identifying the $\Delta E$ as $k_bT$, one obtains the famous expression on the Hawking temperature 
\begin{equation}
\label{ }
T_{H}=\frac{hc^3}{16\pi^2GMk_b}
\end{equation}
where the 4 fundamental constants appear. It is the temperature at which the black hole radiates.

\section{Thermal emission by an aperture}
\begin{figure}
\begin{center}
\includegraphics[width=15cm]{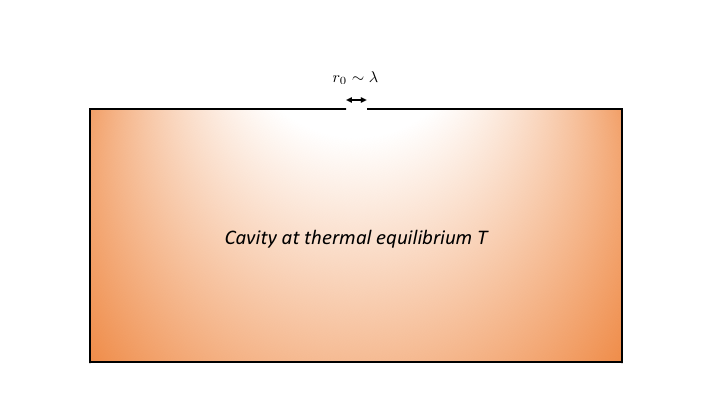}
\caption{Cavity at thermal equilibrium at temperature $T$ separated from the outside by an aperture of radius $r_0$.}
\label{cavity}
\end{center}
\end{figure}
\begin{figure}
\begin{center}
\includegraphics[width=15cm]{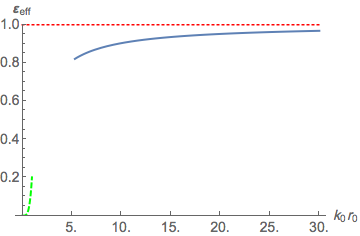}
\caption{Emissivity of the cavity versus $k_0r_0$ calculated by means of fluctuation electrodynamics in the Bethe-Bouwkamp regime ($k_0r_0\ll1$, dashed-green curve) and in the Kirchhoff regime ($k_0r_0>6$, blue curve).}
\label{emissivity}
\end{center}
\end{figure}

In 2016, Joulain et al.\cite{joulain_thermal_2016} studied thermal emission by a subwavelength aperture. Indeed, a small aperture separating a cavity at thermal equilibrium from the outside is often seen as the archetype of a blackbody (Fig. \ref{cavity}). The idea is that such a structure is a perfect absorber since an incident radiation in the cavity will hardly be able to exit from. This remains true as long as the the aperture is large compare to the wavelength involved. When it is not the case, the aperture confines the radiation and diffracts it so that the transmission through the aperture is greatly reduced. This phenomenon is actually nothing else than the usual diffraction by a small aperture. In the case of a cavity which is at thermal equilibrium, the energy density in the cavity $u(\omega)$ is given by the so called Planck formula which is maximum at a frequency given by the Wien's law. 
$$
u(\om)=\frac{\hbar\omega^2}{\pi^2c^3(\exp[\hbar\om/k_bT]-1)}
$$
The angular frequency corresponding to the maximum of emission is given by
$$
\hbar\om_{max}=2.8 k_bT
$$
Therefore, one expects that as long as the aperture is much larger than the wavelength given by the Wien's law, it will behave and emit as a perfect blackbody. The condition reads in the case of an aperture of radius $r_0$
\begin{equation}
\label{ }
r_0\gg\lambda_{max}=2\pi c/\om_{max}
\end{equation}
or if the temperature of the cavity is much larger than a critical temperature
\begin{equation}
\label{ }
T\gg\frac{hc}{2.8r_0k_b}=T_c
\end{equation}
Note that the right hand side of the preceding equation is very similar to the Hawking temperature except that some numerical factors are different and that the Schwarzschild radius is replaced by the aperture radius $r_0$. Joulain et al. have solved the problem using fluctuationnal electrodynamics \cite{rytov_principles_1989} when the aperture is larger that the wavelength ($k_0r_0>6$ with $k_0=\om/c$) or when $k_0r_0\ll1$ (Bethe-Boukamp theory \cite{bethe_theory_1944,bouwkamp_diffraction_1954}). It is possible to choose at what precision an aperture can be assimilated to a blackbody. For example, if one plots the emissivity of such an aperture vs $k_0r_0$ (See Fig.\ref{emissivity}), one shows that this emissivity approaches 1 only for large $k_0r_0$.
What we can emphasis here is that the aperture of a cavity at thermal equilibrium will behave like a blackbody really if its temperature is such that the photons emitted will have their wavelength smaller than the aperture. Similarly, it can be seen that a black hole emits radiation at a temperature which is such that the wavelength associated to this temperature is of the order of the Schwarzschild radius. 

\section{Thermal emission of a small structure}
In the last 20 years, many works have shown that thermal emission could be different whether it was considered in the near or in the far-field i.e at distances small or large compare to the thermal wavelength. For example, the density of electromagnetic energy can be calculated above an interface by means of fluctuationnal electrodynamics. Note that the density of energy is proportionnal to the density of electromagnetic waves times the mean energy of an oscillator at temperature $T$.  When one considers a polar material supporting surface waves such as phonon polaritons, the density of energy increases strongly at subwavelength distances due to the strong increase of the density of states close to the interface. On the contrary, the density of energy at large distance is very close to the one in the vacuum due to the fact that the density of states far from the interface is very close to the one in the vacuum.  However, even if the density of surface modes increases a lot close to the interface, these modes remain confined close to the interface and do not radiate in the far field. Evanescent waves do not contribute to the flux of a flat interface constituted of a single material and do not radiate in the far field. However, it is well known that these waves can be scattered by small objects, coupled to the far-field and can contribute to a strong increase of the radiation emission. This is the principle of scattering of surface waves by a grating \cite{greffet_coherent_2002} or by a microscopic tip \cite{de_wilde_thermal_2006}. The coupling to the far-field can be understood as followed. A surface wave has a parallel wave vector larger than $k_0$ so that the wave vector component perpendicular to the interface is pure imaginary in vacuum, evanescent and cannot propagate. The presence of grating, a tip or a discontinuity couple with the surface wave so that its parallel wave vector can be slightly modified in order that it goes below $k_0$ : the wave can now radiate into the far field. This phenomenon can also happen if the field is confined at a subwavelength distance. 
The analogy with black hole radiation remains here between the radiation of virtual particle-antiparticle pairs and between the scattering of the thermally populated surface waves by subwavelength objects.

In the case of an aperture filled with a material, Joulain et al. \cite{joulain_thermal_2016} gave an expression of the emitted flux which depends on a transmission function of the radiation through the aperture. When the transmission function is 1, one is in the situation of the Kirchhoff approximation so that the heat flux is given by
\begin{eqnarray}
\label{formflux}
\phi(\omega)&=&\phi^0(\omega)\int_0^{2k_0r_0}W(u/k_0)u F(u) du \nonumber\\
&\times&\left\{\int_0^{1}\frac{\kappa J_0(\kappa u) d\kappa}{\sqrt{1-\kappa^2}}(2-|r^s|^2-|r^p|^2)
+\int_{1}^\infty\frac{2\kappa J_0(\kappa u) d\kappa}{\sqrt{\kappa^2-1}}\left[\Im(r^s)+(2\kappa^2-1)\Im(r^p)\right]e^{-2\sqrt{\kappa^2-1}k_0z}\right\}
\end{eqnarray}
where $\kappa=K/k_0$, $F(u)=(\sin u-u\cos u)/u^3$ and $W(d)=\frac{2}{\pi}\left[\arccos\frac{d}{2r_0}-\frac{d}{2r_0}\sqrt{1-\left(\frac{d}{2r_0}\right)^2}\right]$. $W(u/k_0)$ is a function that decreases almost linearly from 1 to 0 when $u$ is increasing from 0 to  $2k_0r_0$ whereas $F(u)$ decreases from 1 to 0 very fast as long as $u$ is larger than 1. Under these conditions, if $k_0r_0$ is much larger than 1, integration over $u$ between 0 and $2k_0r_0$ can be replaced by an integration between 0 and $\infty$. As $F(u)$ decreases fast with $u$ when $u$ is larger than 1, $W$ can be approximated by 1 in the integral. Noting that $\int_0^\infty uF(u)J_0(\kappa u)du$ vanishes if $\kappa>1$ and is equal to  $\sqrt{1-\kappa^2}$ if $\kappa<1$ \cite{gradshteyn_table_nodate}, one retrieves that there is no contribution of the evanescent waves to the emitted flux for large apertures and one retrieves in this case the classical expression of the emitted flux for a large surface
\begin{equation}
\label{ }
\phi=\phi_{clas}=\phi^0(\omega)\int_0^1\kappa d\kappa (2-|r^s|^2-|r^p|^2)=\phi^0(\omega)\int d\Omega\cos\theta\varepsilon(\theta)
\end{equation}
where $\varepsilon(\theta)$ is the material emissivity, $\theta$ the angle with the normal to the surface and $\Omega$ the solid angle over which radiation angular integration is performed. From the preceding, discussion, one sees that the scattering of confined evanescent wave will occur when $k_0r_0$  approaches 1. For thermal radiation, this corresponds to the condition where $r_0\sim\lambda_{max}$ i.e. typical temperature of the order 
\begin{equation}
\label{ }
T\sim\frac{hc}{2.8r_0k_b}
\end{equation}
which is still very similar to the Hawking temperature where the Schwazschild radius has been replaced by the aperture size.

As in the case of vacuum aperture, this critical temperature corresponds to the one associated to the maximum radiation emission wavelength (thermal wavelength) at this temperature which is of the same order of the the object size. 

This temperature also naturally appears when one is dealing with thermal emission of a perfect spherical absorber which size is much smaller than the thermal wavelength.
Let us thus consider a sphere filled with a material which permittivity is $\epsilon$. The scattering cross and absorption cross section can be calculated exactly with Mie Theory. Then, the total emission by the sphere is just the product of the emittance multiplied by the absorption cross section and integrated over all frequencies. 
\begin{equation}
\label{ }
\phi=\int \frac{\hbar\omega^3}{4\pi^2c^2(\exp[\hbar\omega/k_bT]-1)}\sigma_{abs}d\omega
\end{equation}

In the case where the sphere size is much smaller than the thermal wavelength, the absorption cross-section is limited to the 1st term of the Mie Theory. The spherical object is in the dipolar approximation. Following Grigoriev et al. \cite{grigoriev_optimizing_2015}, the spherical object can exhibit a maximum cross section if its permittivity is suitably chosen. Note in that case that the absorption cross section is equal to the scattering cross section 
\begin{equation}
\label{ }
\sigma^{max}_{abs}=\sigma^{max}_{sca}=\frac{5\pi}{k_0^2}
\end{equation}
Integrating over all the frequencies, the maximum flux emitted by a spherical object in the dipolar approximation is
\begin{equation}
\label{ }
\phi^{max}=\frac{5\pi k_b^2T^2}{24\hbar}=1.18\times10^{-12}\ {\rm W}\ {\rm K}^{-2}T^2
\end{equation}
which is apart to a numerical  number the product of the quantum of thermal conductance with the temperature. Note that this flux does not depend on the sphere size (although the optimized dielectric function maximizing absorption cross-section $\sigma_{abs}$ depends on it).
This maximum flux can be compared to the classical emitted flux that would be given in the case of a perfect spherical blackbody of radius $r_0$ with brightness temperature $T_b$
\begin{equation}
\label{ }
\phi^{clas}=4\pi r_0^2\sigma T_b^4
\end{equation}
where $\sigma=\pi^2k_b^4/60\hbar^3c^2$ is the Stefan-Boltzmann constant. The brightness temperature therefore reads $T_b=\sqrt{T_{c2}T}$ where
\begin{equation}
\label{ }
T_{c2}=\sqrt{\frac{25}{32}}\frac{1}{\pi^2}\frac{hc}{r_0k_b}
\end{equation}
Note that this critical temperature is similar to the critical temperature $T_c$ that has been considered previously. Here again, when the temperature is such that the thermal wavelength associated is much smaller than the particle radius, the brightness temperature is larger than the actual temperature and is the geometric average of $T$ and $T_{c2}$.

Finally, one can remember from the preceding considerations, that in analogy with black hole evaporation via the coupling of particle-antiparticle pairs, scattering of thermally populated evanescent waves occur at the surface of an object when their wavelength is of the order or smaller than the object size that is when its temperature is smaller than $T_c$.

\section{Conclusion}
We have shown in this work that a temperature very similar to the one introduced by Stephen Hawking caracterizing the black hole evaporation can be introduced in the case of thermal emission of a small object. When the thermal wavelength associated to thermal emission is smaller than the size of the object, it is known that this thermal emission greatly differs from the one given by usual radiometry i.e geometrical optics. This allows to introduce a critical temperature very similar to the one introduced by Stephen Hawking for black hole radiation. Moreover, when the preceding conditions are reached i.e when the temperature is smaller than the critical temperature, an analogy can be done between the separation of particle-antiparticle pairs due to vacuum fluctuations by a black hole horizon and the scattering of evanescent surface waves due to thermal fluctuations by nano objects.

\end{document}